\documentclass[pra,twocolumn,floatfix]{revtex4}
\usepackage{amsmath,amssymb,amsthm}
\usepackage{graphicx}
\usepackage[caption=false]{subfig}
\usepackage{bbm}
\def\P{\ensuremath{\mathcal{P}}}
\def\id{\ensuremath{\mathbbm{1}}}
\DeclareMathOperator{\Tr}{Tr}
\begin{document}
\title{Full counting statistics of weak measurement}
\author{Antonio \surname{Di Lorenzo}}
\affiliation{Instituto de F\'{\i}sica, Universidade Federal de Uberl\^{a}ndia,\\
 38400-902 Uberl\^{a}ndia, Minas Gerais, Brazil}
\begin{abstract}
A weak measurement consists in coupling a system to a probe in such a way that 
constructive interference generates a large output. So far, only the average output 
of the probe and its variance were studied. Here, the characteristic function for the moments 
of the output is provided. The outputs considered are not limited to the eigenstates 
of the pointer or of its
conjugate variable, so that the results apply to any observable $\Hat{o}$ of the probe. 
Furthermore, 
a family of well behaved complex quantities, the normal weak values, is introduced, 
in terms of which 
the statistics of the weak measurement can be described. It is shown that, within a good approximation, 
the whole statistics of weak measurement is described by a complex parameter, the weak value, 
and a real one. 
\end{abstract}
\maketitle
\section{Introduction}
A weak measurement is but a peculiar interference experiment: 
a quantum system is initially prepared in a superposition of eigenstates of an observable $\Hat{A}$; 
the system interacts with a second quantum system, the probe, prepared in a suitable coherent state, through 
a weak coupling proportional to $\Hat{A}$; as a result, the probe evolves as if subject in parallel to different interactions; the different branches of the probe's wavefunction interfere, as revealed by 
the observation of the pointer variable; as in an ordinary interference experiment  
the fringes disappear if averaged over the whole screen, the same happens with a 
weak measurement; in order to fix a point of the `screen' it is necessary to make a postselection of the system. 
Weak measurement (without postselection) 
was introduced in the pioneering paper of Arthurs and Kelly \cite{Arthurs1965}, 
where it allowed to overcome the quantum mechanical limitation to the joint measurement of 
non-commuting observables. 
The coherence of the detectors manifests in the variance of the outputs \cite{DiLorenzo2011a}, 
resulting in an uncertainty 
relation $\Delta p \Delta q\ge \hbar$, twice as large \cite{Arthurs1988} as the Kennard 
limit \cite{Kennard1927,Robertson1928}. 
In another pioneering paper \cite{Aharonov1988}, Aharonov, Albert, and Vaidman studied 
the average output of a weak measurement of a single observable with post-selection \cite{Ozawa1985}, 
showing that it can be dramatically larger than the maximum eigenvalue. 
This effect was soon attributed to the coherence of the probe \cite{Duck1989}. 
Interestingly, when unpolarized spins are injected into a series of three detectors, 
even though these are initially prepared with no coherence in the readout basis, 
coherence can be induced in the 
middle detector by the interaction with each spin, while the first and third detector provide 
 pre- and post-selection. 
This has observable consequences if the rate at which particles 
are fired into the detectors is higher than the decoherence rate of the latters \cite{DiLorenzo2004}. 
Several experimental realizations of weak measurement have been performed by now \cite{Ritchie1991,Parks1998,Pryde2005,Wang2006,Hosten2008,Dixon2009,Yokota2009,Cho2010,Lundeen2011,
Masataka2011,Kocsis2011}. 
Weak measurement has been contemplated theoretically for applications  
in solid state devices \cite{Williams2008,Romito2008,Shpitalnik2008,Ashhab2009a,Ashhab2009b,Bednorz2010,Zilberberg2011}. 
On a theoretical ground, Ref. \cite{Johansen2004} extended the results of \cite{Aharonov1988} 
to any initial state of the probe; also, the average and the variance of both $\Hat{p}$, 
the pointer variable, and of $\Hat{q}$, its conjugated variable were considered \cite{Jozsa2007}; in Ref. \cite{DiLorenzo2008}, 
the general case of a non-instantaneous interaction was studied, showing that 
the dynamical phase of the detector influences the outcome. 
\begin{figure}[t!]
\includegraphics[height=0.8in]{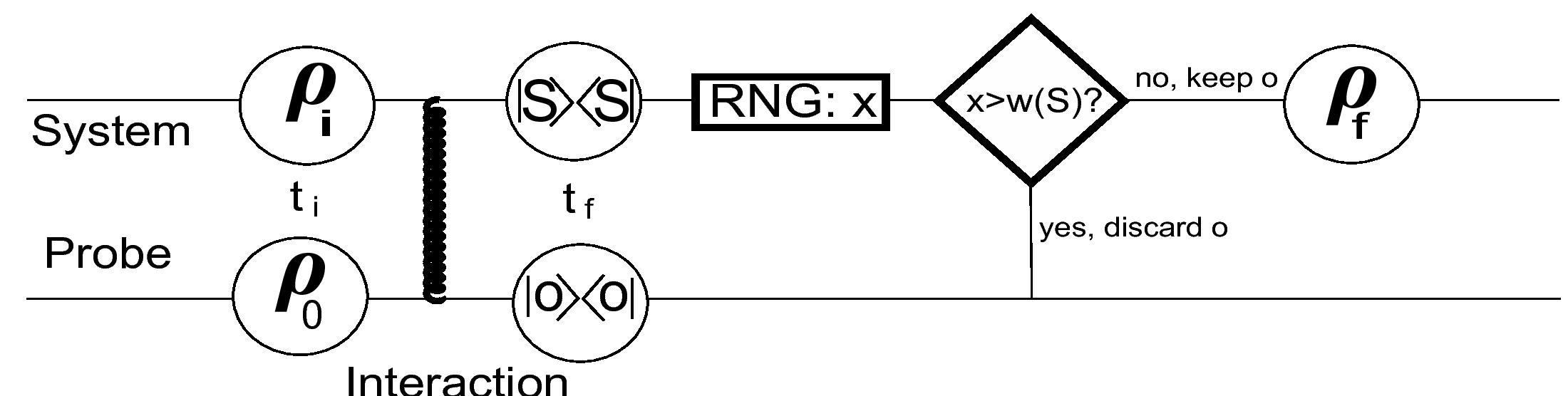}
\caption{
A schematic view of the measurement with pre- and post-selection, the horizontal direction 
representing increasing time. A system is prepared in $\rho_i$, interacts with a probe, then 
is projected into $|S\rangle$ by a measurement. After the interaction, the observable $o$ 
of the probe is measured. 
A random number generator produces a number $0\le x\le 1$. 
If $x>w(S)$, an arbitrarily chosen probability depending on $S$, then the record $o$ 
is discarded.}
\label{fig:scheme}
\end{figure}
Ref. \cite{Aharonov1988} associated a complex quantity, the weak value  $A^w$, 
to the weak measurement. 
Assuming a Gaussian distribution for the probe, the real part of the weak value equals 
the average output. For spin 1/2, this value can easily exceed $\hbar/2$ even for random pre- and 
postselected states \cite{Berry2011}. 
The real part of $A^w$ describes well the average output of the weak measurement  
as far as the pre- and postselected states are not almost orthogonal. 
When they are nearly orthogonal, the output can indeed reach a large value, but the perturbative expansion used in \cite{Aharonov1988} fails, 
for a fixed coupling strenght \footnote{However, given non-orthogonal pre- and post-selected states, one can always choose a coupling weak enough 
that the result of \cite{Aharonov1988} holds. Unfortunately, it is not always possible to tune the interaction strength at will.}, 
since $A^w$ diverges, while the observed 
output has a finite maximum just before the orthogonal configuration, then drops abruptly \cite{DiLorenzo2008,Wu2011}. 
In simple words, weak measurement with postselection relies on a conditional probability which, 
by Bayes' rule \cite{Bayes1763},  
is $\P(o|f)=\P(o,f)/\P(f)$, the ratio of the joint probability $\P(o,f)$ of observing $o$ for the probe and 
making a final postselection $f$,  
over the marginal probability $\P(f)$ of making a successful postselection. 
Both probabilities can be expanded in powers of the interaction strength. For nearly orthogonal 
pre- and postselected state, however, the leading term in the denominator vanishes faster than the 
corresponding term in the numerator for the average $\int\!do\, o \P(o,f)/\P(f)$. 
This artificial divergence should be treated by expanding both 
numerator and denominator up to the first non vanishing terms. This apparently straightforward  
prescription was missed so far by the community working with weak measurement, with the exception of 
\cite{Wu2011}, where the correct expansion, however, was made only for the denominator. 

Furthermore, while the experiments provide the full probability distribution $\P(o|f)$, the theory 
generally considers only average values, and at times variances \cite{Jozsa2007,Parks2011}. 
Here, we provide the full counting statistics, i.e., the characteristic function for all the moments of any observable 
$\Hat{o}$ of the probe, conditioned on the postselection. We give a correct expansion working 
for any pre- and postselected state, and a somewhat simpler interpolating formula working satisfactorily 
over the whole range of postselection. 
\section{Measurement followed by postselection}
We start by describing an ideal, instantaneous (von Neumann) measurement with postselection. 
Let us consider a quantum system prepared, at time $t_i-\varepsilon$, in a 
state $\Hat{\rho}_S(t_i-\varepsilon)=\Hat{\rho}_{i}$ (preselection), with 
$\varepsilon$ an infinitesimal time. 
The system interacts, at time $t_i$, with another quantum system, the probe, 
through 
$H_{int} = -\hbar\lambda \delta(t-t_i) \Hat{q} \Hat{A}$,
where $\Hat{A}$ is an operator on the system's Hilbert space, $\Hat{q}$ on the probe's. 
The probe is prepared 
in a state $\Hat{\rho}_P(t_i-\varepsilon)=\Hat{\rho}_{0}$.  
At time $t_f=t_i+\varepsilon$ a projective measurement 
of the observable $\Hat{o}$ is made on the probe. 
The procedure is particularly interesting, when $\Hat{o}=\Hat{p}$ 
is conjugate to $\Hat{q}$, $[\Hat{q},\Hat{p}]=i$, since then it   
is the observable of the probe that carries information about the measured quantity $\Hat{A}$. 
At time $t_f$, a projective measurement of an observable $\Hat{S}_f$ of the system is made, giving an output $S$ and leaving the system in the state $|\Hat{S}_f\colon S\rangle$. 
Then, given $S$, one keeps the record $o$ according to an arbitrarily 
chosen probability $w(S)$. 
This leaves the system in the postselected mixed state 
$\Hat{\rho}_f=\sum_{S} w(S) |\Hat{S}_f\colon S\rangle\langle\Hat{S}_f\colon S|/W$, with 
$W=\sum_S w(S)$. 
The procedure detailed above describes a measurement with pre- and post-selection and it is sketched 
 in Fig. \ref{fig:scheme}.

Since the interaction is instantaneous, the problem can be treated analytically. 
The joint state for the probe and the system after the interaction is 
\begin{align}
\Hat{\rho}_{P,S} =  
e^{i\lambda \Hat{A} \Hat{q}}
\Hat{\rho}_i\otimes\Hat{\rho}_0 
e^{-i\lambda \Hat{A} \Hat{q}},
\label{eq:finstate0}
\end{align}
or, in the basis of eigenstates of $\Hat{p}$ and $\Hat{A}$, 
\begin{align}
\rho(p,a,p',a';t_f) =& \ 
\rho_{0}(p-\lambda a,p'-\lambda a') \rho_{i}(a,a').
\label{eq:finstate}
\end{align}

The conditional density matrix for the probe given 
an initial preparation on $\Hat{\rho}_0$, a system preselected in $\Hat{\rho}_i$ and 
successfully postselected in $\Hat{\rho}_f$
at time $t_f$, follows readily after applying Born's rule to Eq.~\eqref{eq:finstate0} 
\begin{align}
\Hat{\rho}_{P|0,i,f} = N^{-1}  
\Tr_S{\{\Hat{\rho}_f\otimes \id\, e^{i\lambda \Hat{A} \Hat{q}}\Hat{\rho}_i\otimes\Hat{\rho}_0e^{-i\lambda \Hat{A} \Hat{q}}\}},
\label{eq:condrhop0}
\end{align}
with $\mathrm{Tr}$ the trace, and the subscripts $S$ referring to the system (in the following, 
the subscript $P$ will refer to the probe). 
In the readout basis, 
\begin{align}
\rho(p,p'|0,i,f) =  
\sum_{a,a'}& \rho_{0}(p-\lambda a,p'-\lambda a') G(a,a'), 
\label{eq:condrhop}
\end{align}
where $0,i,f$ indicate the conditions 
(from now on we shall write only $f$ and take $0,i$ as 
implicit conditions), 
and with the definition 
\begin{align}
&G(a,a'):= \frac{1}{N}
\langle a'|\Hat{\rho}_f |a\rangle\langle a|\Hat{\rho}_i|a'\rangle ,
\end{align}
while the normalization is 
\begin{align}
\nonumber
N=\frac{\P(f)}{W}
=&
\Tr_{S,P}{\{\Hat{\rho}_f\otimes \id\, e^{i\lambda \Hat{A} \Hat{q}}\Hat{\rho}_i\otimes\Hat{\rho}_0\, 
e^{-i\lambda \Hat{A} \Hat{q}}\}}
\\
\nonumber
=&\int dp \sum_{a,a'} \rho_{0}(p-\lambda a,p-\lambda a') 
\langle a'|\Hat{\rho}_f |a\rangle\langle a|\Hat{\rho}_i|a'\rangle 
\\
=& \int dq \check{\rho}_0(q,q)
Z^w(\lambda q,\lambda q) 
\label{eq:margprob}
, 
\end{align}
where $Z^w(\mu,\nu)$ is the weak characteristic function, defined by 
\begin{equation}
Z^w(\mu,\nu)\equiv \Tr_S{\{\Hat{\rho}_f e^{i\mu\Hat{A}}\Hat{\rho}_i e^{-i\nu\Hat{A}}\}}.
\end{equation}
$N$ is proportional to $\P(f)$, which is the probability of making a successful postselection in 
$\Hat{\rho}_f$ irrespectively of the value of $p$, or of the fact that $\Hat{p}$, 
or any other observable of the probe, was measured at all. 

The joint probability of making a successful post-selection and 
observing an eigenvalue $o$ of $\Hat{o}$ for the probe is then 
\begin{align}\label{eq:exjprob}
\P(o,f)=&
W\Tr_{S,P}{\{\Hat{\rho}_f\otimes \Hat{\Pi}_o\, e^{i\lambda \Hat{A} \Hat{q}}\Hat{\rho}_i\otimes\Hat{\rho}_0\, 
e^{-i\lambda \Hat{A} \Hat{q}}\}}, 
\end{align}
with $\Hat{\Pi}_o=|o\rangle\langle o|$. The conditional probability is $\P(o|f)=\P(o,f)/\P(f)$. 

Eq.~\eqref{eq:condrhop0} can be conveniently rewritten in the $q$-representation
\begin{align}
\label{eq:condrhoq}
\check{\rho}(q,q'|f) =& N^{-1}  
\check{\rho}_0(q,q')
Z^w(\lambda q,\lambda q') .
\end{align}
Equation \eqref{eq:condrhoq} holds also when $\Hat{q}$ does not have a conjugate 
variable. In this general case $\Hat{q}$ may have a discrete spectrum, and then 
the integrals are to be interpreted as Lebesgue-Stieltjes ones. 
The exact conditional probability for any observable $\Hat{o}$ is then 
\begin{equation}\label{eq:genprob0}
\P(o|f)=\frac{1}{N}\int dq dq' \langle q'|\Hat{\Pi}_o|q\rangle Z^w(\lambda q,\lambda q') 
\check{\rho}_0(q,q'). 
\end{equation}
The characteristic function is obtained by the 
substitution $\exp{[i\chi \Hat{o}]}\to \Hat{\Pi}_o$ in Eq.~\eqref{eq:genprob0}, 
\begin{equation}\label{eq:condgenfunc}
Z(\chi|f)=\frac{1}{N}\int dq dq' \langle q'|e^{i\chi\Hat{o}}|q\rangle Z^w(\lambda q,\lambda q') 
\check{\rho}_0(q,q'). 
\end{equation}

Equation~\eqref{eq:condrhop} shows the peculiarity that the statistics of $p$, 
$\P(p|f)=\rho(p,p|f)$,  
is influenced by the off-diagonal elements of $\rho_0(p,p')$, 
so that interference effects may be relevant. By contrast, Eq.~\eqref{eq:condrhoq} reveals that no interference effects arise if 
one chooses to observe $\Hat{q}$ rather than $\Hat{p}$, so that this case is of lesser relevance. 
Interference terms are appreciable when $\lambda \delta a\ll \delta p$, 
where $\delta a$ is the spacing between eigenvalues of $\Hat{A}$ and 
$\delta p$ is the coherence scale in $p$, i.e., the value over which the 
off-diagonal elements $\rho(p+\delta p,p-\delta p)$ die out. 
A measurement performed in this regime is called weak measurement. 
The state of the system after a weak measurement is almost unaffected, but may acquire 
a small perpendicular component to its initial state. In the rare cases when the postselection is made 
on an almost perpendicular state, so that $\mathrm{Tr}_S\{\Hat{\rho}_f\Hat{\rho}_i\}\simeq 0$, 
the average value of $p$ can be finite, due to the constructive interference represented 
by the contribution of the off-diagonal elements. Since for a strong measurement there is a one-to-one 
correspondence between the final state of the detector and that of the system, so that 
$A=p/\lambda$ is associated to the system, 
if one makes stubbornly the same position, then the weak measurement 
of $\Hat{A}$ can yield a large average output. This is a purely quantum effect, since if it was due simply 
to a large classical uncertainty in the initial distribution $\rho_0(p,p)$, then each individual measurement 
may give a large value, but this would be washed out by averaging.   

\section{Normal weak values}
The results presented so far are exact, but not always susceptible of an easy evaluation. 
In the following, we shall make an expansion in the interaction $\lambda$, assumed to be 
small compared to some appropriate scales that will arise in due course. 
A set of quantities emerging naturally from the expansion are  
the normal weak values, defined as 
\begin{subequations}
\begin{align}
&\alpha_{j,k}^w:=
 \left.\frac{(i\partial)^{j+k} Z^w(\mu,\nu )}{\partial \mu^k(-\partial \nu)^j}
\right|_{\substack{{\mu=0}\\{\nu=0}}}
=\mathrm{Tr}_{S}\{\Hat{A}^j\rho_f \Hat{A}^k\rho_i\} . 
\label{eq:normwv}
\end{align}
We notice that $\alpha_{k,j}^w=\alpha_{j,k}^{w*}$, so that $\alpha_{j,j}^w$ are real. 
For nearly orthogonal pre- and postselected state (NOPPS), $\alpha_{0,0}^w\to 0$ and $\alpha_{j,0}^w\to 0$, in such a way that $\alpha_{0,0}^w/\alpha_{j,0}^w\to 0$ for $j\ge 1$; also,  
$\alpha_{j,k}^w\not\to 0$ for $j,k\ge 1$. 
There are two exceptions to this behaviour: \\
1) Either $\rho_i$ or $\rho_f$ is a mixture of degenerate eigenstates of $\Hat{A}$. 
This case bears no interest, since one is basically repeating the same measurement, 
the second time in the weak regime, the first or the third in the strong one. 
Thus, it is impossible to observe a 
postselected state which is orthogonal to the preselected one, which reflects in
all $\alpha^w_{j,k}$ being zero, so that the conditional 
probability takes the indeterminate form $\tfrac{0}{0}$. \\
2) Neither $\rho_i$ nor $\rho_f$ is a mixture of degenerate eigenstates of $\Hat{A}$, but 
$\Hat{A}$ is $n$-idempotent, 
$\Hat{A}^{n+1}=\Hat{A}$ for some $n\ge 1$ (up to a factor \footnote{It is sufficient that $\Hat{A}^{n+1}=k\Hat{A}$. In this case, one can redefine the operator and the coupling constant $\Hat{A}\to k^{1/n}\Hat{A}$ and $\lambda\to k^{-1/n}\lambda$.}). This implies that $\Hat{A}^{n}=\Hat{A}^{2n}$ is a projection operator (over the subspace spanned by  the eigenvectors of $\Hat{A}$ with non-zero eigenvalue). 
Spin 1/2 and spin 1 (with $n=2$), 
and yes-no observables, i.e., projection operators (with $n=1$), are important instances.  
In this second case, 
$\alpha^w_{j,n}$ vanishes as fast as $\alpha^w_{j,0}$ for NOPPS if either $\rho_i$ or $\rho_f$ 
is a mixture of states belonging to the subspace defined by $\Hat{A}^n$ (which may  
coincide with the whole vector space if $\Hat{A}^n=1$); 
consequently, $\alpha_{0,0}^w/\alpha_{0,kn}^w\to 1$ and $\alpha_{j,kn}^w\to 0$ for such states. 
However, $\alpha_{1,1}^w\not\to 0$ for NOPPS that do not fall under the previous case, 
so that the conditional probability is not indeterminate. 

For brevity we put $\alpha_{j,0}^w=\alpha_j^w$ and in particular $\alpha_1^w=\alpha^w$. 
The normal weak values appear naturally in the expansion of $N$ and $\rho(p,p')$. 
For historical reasons, the canonical weak value used in the literature is $A^w=\alpha_1^w/\alpha_0^w$, 
which diverges for NOPPS. A recent paper \cite{Wu2011} introduced a family of diverging quantities 
$A^w_{j,k}=\alpha^w_{j,k}/\alpha_0^w$ in terms of which to describe the weak measurement. 
As this approach may obscure the relatively simple problem at hand, we shall use the well behaved 
normal weak values. 

The normal weak values (and the canonical weak value) arise naturally 
whenever the initial density matrix for the detector $|\check{\rho}(q,q')|$ 
is peaked around a value $(q^*,q^*)$ such that $\lambda q^*\ll 1$. 
If this is not the case, then one should define 
\begin{align}
\alpha_{j,k}^w:=&
 \left.\frac{(i\partial)^{j+k} Z^w(\mu,\nu )}{\partial \mu^k(-\partial \nu)^j}
\right|_{\substack{{\mu=\lambda q^*}\\{\nu=\lambda q^*}}}
\nonumber
\\
=& \mathrm{Tr}_{S}\{\Hat{A}^j\rho_f \Hat{A}^k 
e^{i\lambda q^*\Hat{A}}\rho_i 
e^{-i\lambda q^*\Hat{A}}\} . 
\label{eq:normwvb}
\end{align}
\end{subequations}
These values are no longer specific to the system, but 
depend also on the coupling and the detector property $q^*$. 

In order to express the characteristic function later on, we define a one-parameter family 
of weak values 
\begin{subequations}
\begin{align}
\alpha_{j,k}^w(z):=&
 \left.\frac{(i\partial)^{j+k} Z^w(\mu,\nu )}{\partial \mu^k(-\partial \nu)^j}
\right|_{\substack{{\mu=z}\\{\nu=-z}}}
\nonumber
\\
=&\mathrm{Tr}_{S}\{\Hat{A}^j\rho_f \Hat{A}^k
e^{iz\Hat{A}}\rho_i e^{iz\Hat{A}}\} , 
\label{eq:normwvchi}
\end{align}
or, if $\lambda q^*$ is appreciable,  
\begin{align}
&\alpha_{j,k}^w(z):=
 \left.\frac{(i\partial)^{j+k} Z^w(\mu,\nu )}{\partial \mu^k(-\partial \nu)^j}
\right|_{\substack{{\mu=z+\lambda q^*}\\{\nu=-z+\lambda q^*}}}
\nonumber
\\
=&
\mathrm{Tr}_{S}\{\Hat{A}^j\rho_f \Hat{A}^k
e^{i(z+\lambda q^*)\Hat{A}}\rho_i e^{i(z-\lambda q^*)\Hat{A}}\} . 
\label{eq:normwvchib}
\end{align}
\end{subequations}
\section{Characteristic function for $q$}
We proceed to evaluate the full counting statistics of 
$\check{\P}(q|f)$, i.e., its Fourier transform 
$\check{Z}_Q(\chi)=\int dq \exp{[i\chi q]} \check{\P}(q|f)$. 
We need first to expand $N$, which from Eq.~\eqref{eq:margprob}
 (see also, e.g., \cite{Wu2011}) results  
\begin{align}
N=& \sum_{n=0}  \frac{\overline{(-i\lambda q)^n}}{n!} 
\sum_{j=0}^{n} (-1)^j \binom{n}{j} \alpha_{j,n-j}^w ,
\end{align}
where the bar denotes average with $\Hat{\rho}_0$. 
Then, after expanding $Z^w$ in Eq.~\eqref{eq:condgenfunc}, we have 
\begin{equation}
\label{eq:chiq}
\check{Z}_Q(\chi)= N^{-1}{\sum_{n=0} \frac{\overline{(-i\lambda q)^n e^{i\chi q}}}{n!}  
\sum_{j=0}^n (-)^j \binom{n}{j} \alpha_{j,n-j}^w
}.
\end{equation}
In order to have an expansion working for NOPPS, we expand both the numerator 
and denominator to second order, which contains the non-vanishing quantities 
$\alpha_{1,1}^w$ (notice that Ref. \cite{Wu2011}, when estimating the average, 
inconsistenly expands the numerator up to the first order and the denominator to the second order, so that the case of orthogonal pre- and postselected states has to 
be treated as a separate case, unnecessarily) 
\begin{align}
\nonumber
Z_Q(\chi)=& N_2^{-1}\biggl\{\overline{e^{i\chi q}} \alpha_0^w+2\lambda \overline{q e^{i\chi q}}
\mathrm{Im}(\alpha_{1}^w)\\
\label{eq:chiqappr}
&+
\lambda^2 \overline{q^2 e^{i\chi q}}
\left(\alpha_{1,1}^w-\mathrm{Re}(\alpha_{2}^w) \right)
\biggr\} , 
\end{align}
with $\overline{\exp{[i\chi\Hat{q}]}}$ the initial characteristic function  
and 
\begin{equation}
\label{eq:den2}
N_2=\alpha_0^w+2\lambda\overline{q} \mathrm{Im}(\alpha_1^w)+\lambda^2 \overline{q^2}
\left[\alpha_{1,1}^w-\mathrm{Re}(\alpha_{2}^w)\right]  .
\end{equation}

After considering that higher order terms in $\lambda$ become important only for NOPPS, 
so that $\alpha_{1,1}^w\gg \alpha_2^w$, we can use a simpler interpolation formula within 
a satisfactory approximation
\begin{align}
\nonumber
Z_Q(\chi)= {N'_2}^{-1}\biggl\{&\overline{e^{i\chi q}}\alpha_0^w+2\lambda \overline{q e^{i\chi q}}
\mathrm{Im}(\alpha_{1}^w)\\
&+
\lambda^2 \overline{q^2 e^{i\chi q}}
\alpha_{1,1}^w 
\biggr\} , 
\label{eq:chiqappr2}
\end{align}
with 
\begin{equation}
\label{eq:den3}
N_2'=\alpha_0^w+2\lambda\bar{q} \mathrm{Im}(\alpha_1^w)+\lambda^2 \overline{q^2}
\alpha_{1,1}^w .
\end{equation}
However, if one is deriving the $j$-th moment $\langle q^j\rangle_f$ from $Z_Q$, one 
should keep the term $\alpha_2^w$ in the numerator of Eq.~\eqref{eq:chiqappr} 
whenever $\overline{q^{j+1}}=0$  
or $\overline{q^{j+1}}\approx \lambda \overline{q^{j+2}}$, since then 
the first order contribution vanishes or it is comparable to the second order one.

In all the equations, one may divide both numerator and denominator by $\alpha_0^w$, 
so that the simpler interpolation formula is a function of a complex number, the canonical 
weak value, 
$A^w=\Tr{\{\Hat{\rho}_f\Hat{A}\Hat{\rho}_i\}}/\Tr{\{\Hat{\rho}_f\Hat{\rho}_i\}}$, 
and a real number $B^w=\Tr{\{\Hat{\rho}_f\Hat{A}\Hat{\rho}_i\Hat{A}\}}/\Tr{\{\Hat{\rho}_f\Hat{\rho}_i\}}$. 
The inequality $B^w\ge |A^w|^2$ holds, with the equality sign for pure pre- and 
postselected states. 
For NOPPS $B^w$ diverges faster than $A^w$, dominating the expansion. 
When, as discussed above, $\overline{q^{j+1}}=0$  
or $\overline{q^{j+1}}\approx \lambda \overline{q^{j+2}}$, an additional complex parameter should be used 
$C^w=\Tr{\{\Hat{\rho}_f\Hat{A}^2\Hat{\rho}_i\}}/\Tr{\{\Hat{\rho}_f\Hat{\rho}_i\}}$ in order to get a 
good interpolation for $\langle q^j\rangle_f$. 

\section{Characteristic function for the readout}
\subsection{Special case: Gaussian probe}
Let us say that the probe is described by 
an initial Wigner function 
\begin{equation}
W_0(Q,P)\!=\!\frac{1}{2\pi\Delta P\Delta Q}
\exp\!{\left[-\frac{P^2}{2\Delta P^2} -\frac{(Q-\bar{q})^2}{2\Delta Q^2}\right]} . 
\end{equation}
We recall that the coherence scale $\delta p$ in the readout basis is 
$\delta p=1/2\Delta Q$.  
The exact expression for the probability of observing $p$ is 
\begin{widetext}
\begin{equation}
\P(p|f)=\frac{1}{\sqrt{2\pi}\Delta P} 
\sum_{a,a'} \exp{\left[i\lambda \bar{q}(a-a')-\frac{\lambda^2 \Delta Q^2}{2} (a-a')^2\right]} \exp{\left\{-\frac{[p-\lambda(a+a')/2]^2}{2\Delta P^2}\right\}}
G(a,a')
. 
\end{equation}
\end{widetext}
While one can expand  the first exponential in the numerator 
in $\lambda \bar{q}$, $\lambda \Delta Q$, provided these are small, 
it would be a grave error 
to expand the exponential containing $p$ in $\lambda/\Delta P$, as 
it is appreciable only when $p$ takes values 
order $\lambda$. 
In order to calculate the moments of $p$, $\langle p^j\rangle_f=\int dp \P(p|f) p^j$one 
should change the integration variable to $p'=p-\lambda(a+a')/2$, so that 
$p^j=[p'+\lambda(a+a')/2]^j$, and expand the binomial. 
By following the above prescription, and keeping terms up to $\lambda^2$,  
we obtain, for even $j$, 
\begin{align}
\langle p^j\rangle_f=& \overline{p^j}+ 
\frac{\lambda^2}{2}
\binom{j}{2} \overline{p^{j-2}}  
\left[\alpha^w_{1,1}+\mathrm{Re}(\alpha^w_2)\right]
/N_2 ,
\label{eq:gaussmomseven}
\end{align}
and, for odd $j$, 
\begin{align}
\langle p^j\rangle_f=&   \lambda\left[
j \overline{p^{j-1}}\mathrm{Re}(\alpha^w_1)
+{\lambda}j\overline{q} \overline{p^{j-1}} \mathrm{Im}(\alpha^w_2) \right]/N_2, 
\label{eq:gaussmomsodd}
\end{align}
The bar means average with respect to $\P_0(p)$. 
After considering that higher order terms in $\lambda$ become important only for NOPPS, 
so that $\alpha_{1,1}^w\gg \alpha_2^w$, we can use within 
a satisfactory approximation
\begin{equation}
\label{eq:den2p}
N_2'=\alpha_0^w+2\lambda\overline{q} \mathrm{Im}(\alpha_1^w)+\lambda^2 \overline{q^2}
\alpha_{1,1}^w
.
\end{equation}
We may now divide both numerator and denominator by $\alpha^w_0$, 
in order to eliminate one parameter, as done in the previous section, obtaining 
for even $j$, 
\begin{align}
\langle p^j\rangle_f=& \overline{p^j}+ 
\frac{\lambda^2}{2}
\frac{\binom{j}{2} \overline{p^{j-2}}  
\left[B^w+\mathrm{Re}(C^w)\right]}
{1+2\lambda\overline{q} \mathrm{Im}(A^w)+\lambda^2 \overline{q^2}
\left[B^w-\mathrm{Re}(C^w)\right] },
\end{align}
and, for odd $j$, 
\begin{align}
\langle p^j\rangle_f=&   \lambda j\frac{
\overline{p^{j-1}}\mathrm{Re}(A^w)
+{\lambda}\overline{q} \overline{p^{j-1}} \mathrm{Im}(C^w)}{1+2\lambda\overline{q} \mathrm{Im}(A^w)+\lambda^2 \overline{q^2}
\left[B^w-\mathrm{Re}(C^w)\right] }, 
\end{align}

The characteristic function, defined as $Z_P(\chi|f)=\langle e^{i\chi p}\rangle_f$, is 
\begin{align}
Z_P(\chi|f)=&
\overline{e^{i\chi p}} Z_S(\lambda\chi)
\label{eq:gausscf0}
\end{align}
with $\overline{e^{i\chi p}}=e^{-\chi^2\Delta P^2/2}$, and 
\begin{equation}
Z_S(\lambda\chi)=
\sum_{a,a'} e^{i\lambda [\chi (a+a')/2+\bar{q}(a-a')]-\frac{\lambda^2 \Delta Q^2}{2} (a-a')^2} G(a,a')
\end{equation}
In particular we have that the $n$-th cumulants (logarithmic derivatives of $Z$) are of 
order $\lambda^n$ for $n\ge 3$. 
After expanding in $\lambda \bar{q}$, $\lambda \Delta Q$	, 
\begin{align}
Z_P(\chi|f)=&
\frac{\overline{e^{i\chi p}}}{N_2} \biggl\{
\alpha_0^w\left(\frac{\lambda\chi}{2}\right) +
2\lambda
\overline{q} \mathrm{Im}\left(\alpha_1^w\left(\frac{\lambda\chi}{2}\right)\right)
\nonumber
\\
&+\lambda^2 \overline{q^2}
\left[\alpha_{1,1}^w\left(\frac{\lambda\chi}{2}\right)-\mathrm{Re}\left(\alpha_2^w\left(\frac{\lambda\chi}{2}\right)\right)\right]
\biggr\},
\label{eq:gausscf}
\end{align}
and, up to second order terms in $\lambda$, 
\begin{align}
Z_P(\chi)=& \frac{\overline{e^{i\chi p}}}{N_2}
\biggl\{ 
\alpha_0^w
+
\lambda\left[i\chi \mathrm{Re}(\alpha_1^w)
+2\overline{q} \mathrm{Im}(\alpha_1^w)\right]
\nonumber
\\
&+\lambda^2 \biggl[
\left(\overline{q^2}-\frac{\chi^2}{4}\right) \alpha_{1,1}^w
+i\overline{q}\chi \mathrm{Im}\left(\alpha_2^w\right)
\nonumber
\\
&\quad \ -
\left(\overline{q^2}+\frac{\chi^2}{4}\right)
\mathrm{Re}\left(\alpha_2^w\right)\biggr]
\biggr\}.
\label{eq:gausscf2}
\end{align}
One should take Eqs. \eqref{eq:gaussmomseven}, \eqref{eq:gaussmomsodd}, and \eqref{eq:gausscf2} 
with a grain of salt: 
in Eqs. \eqref{eq:gaussmomseven} and  \eqref{eq:gaussmomsodd}, 
we have neglected terms of order 
$\binom{n}{k} \lambda^{k} \overline{p^{n-k}}(a+a')^k/2^k$ 
compared to $\binom{n}{2}\lambda^{2} \overline{p^{n-2}} (a+a')^2/4$ for $k>2$. 
Without loss of generality, if $\Hat{A}$ has a limited spectrum, we can rescale 
$\lambda\to \lambda \max{|a|}$, $\Hat{A}\to \Hat{A}/\max{|a|}$, so that 
$|a+a'|/2\le 1$.  
For large enough $n$ the approximation fails, since 
\[\lim_{n\to \infty}\frac{\binom{n}{2} \overline{p^{n-2}}}{\binom{n}{4} \overline{p^{n-4}}}= 0 .\]
Considering that 
$\binom{n}{k} \overline{(p/\Delta P)^{n-k}} x^k$ has a maximum for $k\sim \sqrt{n x^2}$, we have that 
the validity of Eqs. \eqref{eq:gaussmomseven} and \eqref{eq:gaussmomsodd} extends to 
$n^*\simeq \Delta P^2/\lambda^2$. 
The characteristic function 
in Eq.~\eqref{eq:gausscf2}
should be interpreted as giving the correct moments up to $n^*$. 
Since all the moments of $p$ are necessary to reconstruct the probability distribution, 
one would not get the correct $\P(p|f)$ by Fourier transforming $Z(\chi|f)$. 
On the other hand, Eq.~\eqref{eq:gausscf} provides a good approximation to the moments of 
all orders, even though the expansion in $\lambda$ is not done consistently. 
Once Eq.~\eqref{eq:gausscf} is Fourier-transformed to the conditional probability  
$\P(p|f)$, it captures the shifts $p\to p-\lambda a$ induced by the interaction, 
contrary to Eq.~\eqref{eq:gausscf2}. 
\subsection{General case}
The former particular case suggests the proper treatment of the expansion in $\lambda$. 
We define the function 
\begin{equation}
R_0(P,p):=\rho_0(P+p/2,P-p/2) \ .
\end{equation}
Substituting in Eq.~\eqref{eq:condrhop} 
for $p'=p$, we have the conditional probability 
\begin{align}
\P(p|f) = 	  
\sum_{a,a'}& R_{0}(p-\lambda \frac{a+a'}{2},\lambda (a'-a)) 
G(a,a')
.
\label{eq:condrhop2}
\end{align}
The correct way to proceed is to expand $R_0$ in its second argument only. 
For the Gaussian distribution of the preceding subsection, 
$R_0$ factors into the product of a function of $(p+p')/2$ 
and one of $p-p'$. This is not generally the case. 
For the characteristic function, we have 
\begin{align}
Z_P(\chi|f) =& 
\sum_{a,a'} \int dp e^{i\chi p} R_{0}\left(p-\lambda \frac{a+a'}{2},\lambda (a'-a)\right) 
 G(a,a')
\nonumber
\\
\simeq & N_2^{-1}
\biggl\{
\overline{e^{i\chi p}}\alpha_0^w\left(\frac{\lambda\chi}{2}\right) 
+
2\lambda
\overline{e^{i\chi p}q}\ \mathrm{Im}\left(\alpha_1^w\left(\frac{\lambda\chi}{2}\right)\right)
\nonumber
\\
&+\lambda^2 \overline{e^{i\chi p}q^2}
\left[\alpha_{1,1}^w\left(\frac{\lambda\chi}{2}\right)
-\mathrm{Re}\left(\alpha_2^w\left(\frac{\lambda\chi}{2}\right)\right)\right]
\biggr\},
\end{align}
where the bar denotes quasi-averages, 
$\overline{f(p) g(q)}=\int dp dq\, f(p) g(q) W_0(q,p)$. 
In terms of properly defined averages, 
\begin{equation}
\overline{f(p) q^j}=\frac{1}{2^{j}}\sum_{k=0}^j \binom{j}{k} 
\Tr_P\{\Hat{q}^{j-k}f\left(\Hat{p}\right)\Hat{q}^k\Hat{\rho}_0\} .
\end{equation}
We notice that whenever $f(p)=const.$ or $g(q)=const.$, 
the quasi-averages coincide with proper 
averages, so that there is no notation clash with the former section and subsection.  
The moments, to order $\lambda^2$, are 
\begin{align}
\langle p^j\rangle_f=& \overline{p^j}+ 
\lambda\left[
j \overline{p^{j-1}}\mathrm{Re}(\alpha^w_1)+
\left(\overline{p^j q}-\overline{p^j} \overline{q}\right)\mathrm{Im}(\alpha^w_1)
\right]
\nonumber \\
&+
\lambda^2\biggl[
\left(\overline{p^j q^2}-\overline{p^j}\ \overline{q^2}\right)
\left[\alpha^w_{1,1}-\mathrm{Re}(\alpha^w_2)\right]
\nonumber 
\\
&\quad \quad +\frac{j(j-1)}{4} \overline{p^{j-2}}  
\left[\alpha^w_{1,1}+\mathrm{Re}(\alpha^w_2)\right]
\nonumber 
\\
&\quad \quad +j\overline{p^{j-1}q}\mathrm{Im}(\alpha^w_2) 
\biggr], 
\label{eq:genmoms}
\end{align}
while the characteristic function can be approximated, with the same proviso as in the former subsection, as 
\begin{align}
Z_P(\chi|f) 
\simeq & N_2^{-1}\biggl\{
\overline{e^{i\chi p}}  \alpha_0^w
\nonumber
\\
&+
\lambda\left[i\chi \overline{e^{i\chi p}}\mathrm{Re}\left(\alpha_1^w \right)+
\overline{q e^{i\chi p}}\ \mathrm{Im}\left(\alpha_1^w\right)
\right]
\nonumber
\\
&+\lambda^2 \biggl[
\left(\overline{q^2 e^{i\chi p}}-\frac{\chi^2}{4}\overline{e^{i\chi p}}\right) \alpha_{1,1}^w
+i\chi\overline{q e^{i\chi p}}\mathrm{Im}\left(\alpha_2^w\right)
\nonumber
\\
&-
\left(\overline{q^2 e^{i\chi p}}+\frac{\chi^2}{4}\overline{e^{i\chi p}}\right)
\mathrm{Re}\left(\alpha_2^w\right)\biggr]
\biggr\},
\label{eq:chip}
\end{align}
As for the Gaussian case, the validity of Eq.~\eqref{eq:genmoms} is limited to $j\le n^*$. 
The estimation of $n^*$ depends on the actual form of the initial density matrix of the probe, and now it relies on comparing 
$\binom{j}{k}\overline{p^{j-k}}$, $\binom{j}{k}\overline{q p^{j-k}}$, and 
$\binom{j}{k}\overline{q^2 p^{j-k}}$ to, respectively, $\binom{j}{2}\overline{p^{j-2}}$, $\binom{j}{2}\overline{qp^{j-2}}$, and $\binom{j}{2}\overline{q^2p^{j-2}}$. 
\section{Characteristic function for any observable}
\subsection{Small $\lambda q^*$}
Finally, the results are easily extended to the case when one chooses to measure 
a generic observable $\Hat{o}$ of the detector, instead of $\Hat{p}$ or $\Hat{q}$. 
After expanding Eq.\eqref{eq:condgenfunc}, 
assuming that $\lambda q^*\ll 1$ (we recall that $q^*$ indicates the maximum of 
the distribution $\rho_0(q,q)$), 
we have 
\begin{align}
Z_O(\chi|f)=& N^{-1}\sum_{n=0} \frac{(-i\lambda)^{n}}{n!}  
\nonumber
\\
&
\times \sum_{j=0}^n (-1)^j \binom{n}{j} \langle\Hat{q}^{n-j}e^{i\chi\Hat{o}}\Hat{q}^j\rangle_0 \alpha_{j,n-j}^w ,
\label{eq:chio}
\end{align}
with $\langle X\rangle_0=\Tr_P{\{X\rho_0\}}$. 
Expanding up to second order terms, 
\begin{align}
\nonumber
Z_O(\chi|f)\simeq N_2^{-1}\biggl\{& 
\langle e^{i\chi \Hat{o}}\rangle_0 \alpha_0^w+\lambda \biggl[
-i\langle [\Hat{q},e^{i\chi \Hat{o}}]\rangle_0\mathrm{Re}(\alpha_{1}^w)\\
\nonumber&
+
\langle \{\Hat{q},e^{i\chi \Hat{o}}\}\rangle_0
\mathrm{Im}(\alpha_{1}^w)\biggr] 
\\
\nonumber
&+\frac{\lambda^2}{2}
\biggl[2\langle\Hat{q}e^{i\chi\Hat{o}}\Hat{q}\rangle_0 \alpha_{1,1}^w
- \langle\{\Hat{q}^2,e^{i\chi\Hat{o}}\}\rangle_0
\mathrm{Re}(\alpha_{2}^w)\\
&\quad \quad -i \langle[\Hat{q}^2,e^{i\chi\Hat{o}}]\rangle_0
\mathrm{Im}(\alpha_{2}^w)
\biggr]
\biggr\},
\label{eq:chio2}
\end{align}
Putting $\Hat{o}=\Hat{q}$ or $\Hat{o}=\Hat{p}$ in 
Eq.~\eqref{eq:chio2}, we recover 
Eqs. \eqref{eq:chiq} and 
\eqref{eq:chip}.
We notice 
that Eq.~\eqref{eq:chio2} should be interpreted 
as generating formulas for the moments $\langle \Hat{o}^j\rangle_f$, 
up to a critical $n^*$ that must be determined case by case according 
to $\Hat{o}$ and $\Hat{\rho}_0$. 
 
As an example, we have that the expectation of $\Hat{o}$ is 
\begin{align}
\nonumber
\langle o\rangle= \frac{1}{N_2}\biggl\{&
\overline{o} \alpha_0^w +\lambda
\left[-i\overline{[\Hat{q},\Hat{o}]}\mathrm{Re}(\alpha_{1}^w)
+\overline{\{\Hat{q},\Hat{o}\}}\mathrm{Im}(\alpha_{1}^w)\right]\\
\label{eq:avo2}
+&\frac{\lambda^2}{2}\!
\left[2\overline{\Hat{q}\Hat{o}\Hat{q}}\alpha_{1,1}^w
-\overline{\{\Hat{q}^2,\Hat{o}\}}\mathrm{Re}(\alpha_{2}^w)
-\overline{i[\Hat{q}^2,\Hat{o}]}\mathrm{Im}(\alpha_{2}^w)
\right]
\biggr\}.
\end{align}
If $\overline{\Hat{q}\Hat{o}\Hat{q}}\neq 0$ one can neglect the terms $\alpha_2^w$ in 
the second order, giving  
\begin{align}
\nonumber
\langle o\rangle= \frac{1}{N''_2}\biggl\{&
\overline{o}  +\lambda
\left[-i\overline{[\Hat{q},\Hat{o}]}\mathrm{Re}(A^w)
+\overline{\{\Hat{q},\Hat{o}\}}\mathrm{Im}(A^w)\right]\\
\label{eq:avob}
+&\lambda^2\overline{\Hat{q}\Hat{o}\Hat{q}}B^w
\biggr\},
\end{align}
where 
\[N_2"=1+2\lambda\overline{q} \mathrm{Im}(A^w)+\lambda^2 \overline{q^2}B^w.\]
 If $\overline{\Hat{q}\Hat{o}\Hat{q}}=0$, the terms $\overline{\{\Hat{q}^2,\Hat{o}\}}\mathrm{Re}(C^w)$
and $\overline{i[\Hat{q}^2,\Hat{o}]}\mathrm{Im}(C^w)$ 
should be retained. If by chance one has also that 
$\overline{\{\Hat{q}^2,\Hat{o}\}}=\overline{i[\Hat{q}^2,\Hat{o}]}=0$, 
one should in principle retain terms up to $\lambda^3$, 
or higher,  in the numerator. However, since by hypothesis $\lambda$ is small compared to the 
other relevant scales, one can neglect altogether these fine corrections, which lead to a small 
deviation of $\langle o\rangle$ from zero for exactly orthogonal pre- and post-selected states. 
 
Finally, for NOPPS, when $\lambda^2 \overline{q^2} B^w\gg \lambda \overline{q}| A^w|$, 
the leading term in the statistics is 
\begin{equation}
Z_O^{orth}(\chi)\simeq \frac{\overline{\Hat{q}e^{i\chi \Hat{o}}\hat{q}}}{\overline{\Hat{q}^2}}.
\end{equation}
This is a universal value, independent of the observable $\Hat{A}$. 
\begin{figure}[t!]
\centering{
\subfloat[]{\includegraphics[width=2.4in]{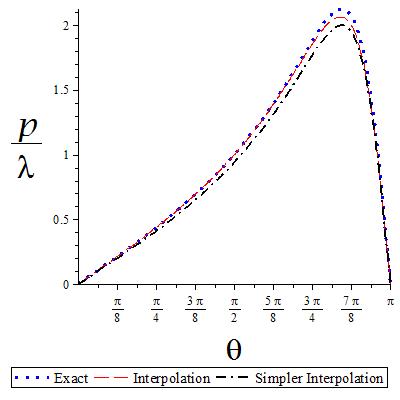}}\\
\subfloat[]{\includegraphics[width=2.4in]{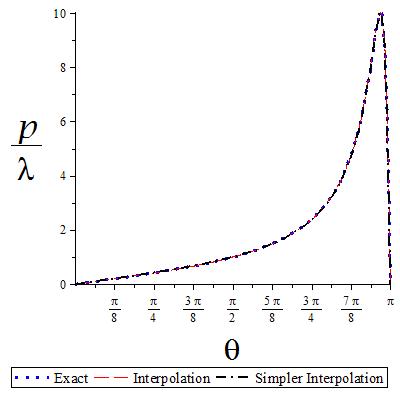}}}
\caption{
\label{fig:comp}
Comparison between the interpolating formula Eq.~\eqref{eq:avo2}, the 
simpler Eq.~\eqref{eq:avob}, and the exact result for the output $p$ of a probe weakly 
measuring a spin component $\mathbf{n}\cdot\boldsymbol{\sigma}$, 
as a function of the angle between the preselected state $|\mathbf{n}_i\!\!\!:\uparrow\rangle$ and the 
postselected one \mbox{$|\mathbf{n}_f\!\!:\uparrow\rangle$}, $\mathbf{n}_f\cdot\mathbf{n}_i=\cos{\theta}$ for a fixed preselection, with 
$\mathbf{n}\cdot\mathbf{n}_i=0$ and all vectors coplanar. 
The figures correspond to different interaction strengths: (a) $\lambda/\delta p=0.5$, (b) $\lambda/\delta p=0.1$.}
\end{figure}
\subsection{Large $\lambda q^*$}
When the term $\lambda q^*$ is large, with $q^*$ identifying a maximum of 
the initial probe distribution $|\check{\rho}_0(q,q')|$ (typically, the initial average value 
$\bar{q}$ should be comparable to $q^*$), the formulas given in the previous three sections 
the weak values must be substituted  with the expressions of 
Eqs.~\eqref{eq:normwvb} and \eqref{eq:normwvchib}, while $q\to q-q^*$. 

In particular, for an initial gaussian state of the detector, $q^*=\overline{q}$, so that 
the even moments of $p$ are 
\begin{align}
\langle p^j\rangle_f=& \overline{p^j}+ 
\frac{\lambda^2}{2}
\frac{\binom{j}{2} \overline{p^{j-2}}  
\left[B^w+\mathrm{Re}(C^w)\right]}
{1+\lambda^2 \overline{(q-\bar{q})^2}
\left[B^w-\mathrm{Re}(C^w)\right] },
\end{align}
and the odd ones 
\begin{align}
\langle p^j\rangle_f=&   \lambda j\frac{
\overline{p^{j-1}}\mathrm{Re}(A^w)}{1+\lambda^2 \overline{(q-\bar{q})^2}
\left[B^w-\mathrm{Re}(C^w)\right] }.
\end{align}

\section{Example: application to a spin 1/2 or q-bit}
We apply the results to the weak measurement of a spin component $\Hat{A}=\mathbf{n}\cdot\boldsymbol{\sigma}$, 
for which 
there is an analytic solution \cite{DiLorenzo2008}. 
The exact conditional probability for the readout variable $p$ is 
\begin{align}
\nonumber
\P(p|f)=&\frac{1}{4N} \sum_{\sigma} 
\left[\alpha_0^w+\alpha_{1,1}^w+2\sigma \mathrm{Re}(\alpha_1^w)\right] 
\rho_0(p-\lambda\sigma,p-\lambda\sigma)
\\
&+
\left[\alpha_0^w-\alpha_{1,1}^w+2i\sigma \mathrm{Im}(\alpha_1^w)\right] 
\rho_0(p+\lambda\sigma,p-\lambda\sigma), 
\label{eq:genspinprob}
\end{align}
the normalization being 
\begin{align}
\nonumber
N=&\frac{1+\overline{\cos{(2\lambda q)}}}{2}\alpha_0^w
+\overline{\sin{(2\lambda q)}}\mathrm{Im}(\alpha_1^w)
\\
\label{eq:genspinnorm}
&+\frac{1-\overline{\cos{(2\lambda q)}}}{2}\alpha_{1,1}^w .
\end{align}
The normal weak values are 
\begin{subequations}
\begin{align}
\alpha_0^w=& \frac{1}{2}\left(1+\mathbf{n}_i\cdot\mathbf{n}_f\right),\\
\alpha_1^w=& 
\frac{1}{2}\mathbf{n}\cdot\left(\mathbf{n}_i+\mathbf{n}_f+i \mathbf{n}_i\times\mathbf{n}_f\right),\\
\alpha_{1,1}^w=&\frac{1}{2} \left(1-\mathbf{n}_i\cdot\mathbf{n}_f+2
\mathbf{n}\cdot\mathbf{n}_i \ \mathbf{n}\cdot\mathbf{n}_f\right),
\end{align}
\end{subequations}
where $\mathbf{n}_i$, $\mathbf{n}_f$ are the polarizations of the pre- and post-selected density matrix, 
i.e., $\rho_{i,f}=(1/2)[1+\mathbf{n}_{i,f}\cdot\boldsymbol{\sigma}]$. 
Equations \eqref{eq:genspinprob} and \eqref{eq:genspinnorm} are new general results, 
since so far in the literature only an initial Gaussian state was considered for the probe \cite{DiLorenzo2008,Geszti2010}. 
\begin{figure}[t!]
\centering{
\subfloat[]{\includegraphics[width=2.4in]{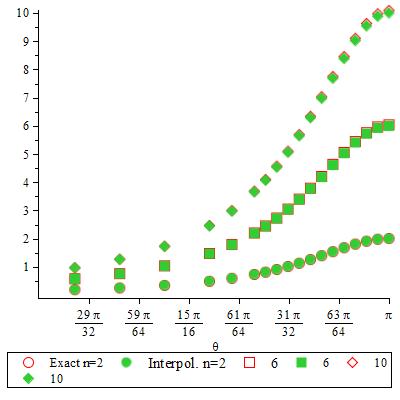}}\\
\subfloat[]{\includegraphics[width=2.4in]{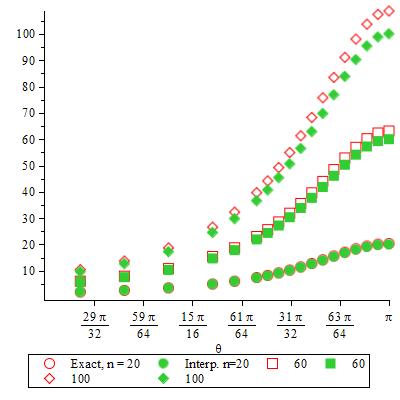}}}
\caption{
\label{fig:compmoms}
Comparison between the interpolating formula (solid symbols) Eq.~\eqref{eq:gaussmomseven}, and the exact result (unfilled symbols) Eq.~\eqref{eq:exmomseven} for the output 
$\langle p^n\rangle_f-\overline{p^n}$ in units 
of $\overline{p^n}$ as a function of the angle between the preselected state $|\mathbf{n}_i\!\!\!:\uparrow\rangle$ and the 
postselected one \mbox{$|\mathbf{n}_f\!\!:\uparrow\rangle$}, $\mathbf{n}_f\cdot\mathbf{n}_i=\cos{\theta}$ for a fixed preselection, with 
$\mathbf{n}\cdot\mathbf{n}_i=0$ and all vectors coplanar. 
Interaction strength:  $\lambda/\delta p=\lambda/\Delta P=0.1$. 
For the values of $\theta$ not shown the discrepancy is undiscernible to the naked eye.}
\end{figure}

For definiteness, we take the probe to be initially described by the Wigner function 
$W_0(p,q)\propto\exp{\{-p^2/2\Delta P^2-2\delta p^2 (q-\bar{q})^2\}}$, where $\delta p\le \Delta P$ 
represents its coherence scale. Then, 
the exact probability distribution for the rescaled readout $A=p/\lambda$ is 
\begin{align}
\nonumber
\P(A|f)=&\frac{1}{4\sqrt{2\pi}N\Delta} \\
\nonumber
&\sum_{\sigma}\biggl\{ 
\left[\alpha_0^w+\alpha_{1,1}^w+2\sigma \mathrm{Re}(\alpha_1^w)\right] e^{- (A-\sigma)^2/2\Delta^2} \\
&+ \left[\cos{(2\lambda\bar{q})}\left(\alpha_0^w-\alpha_{1,1}^w\right) 
+2\sin{(2\lambda\bar{q})}\mathrm{Im}(\alpha_1^w)\right]
\nonumber
\\
&\times e^{- A^2/2\Delta^2-1/2\delta^2} \biggr\} ,
\end{align}
with $\Delta \equiv \Delta P/\lambda$, $\delta \equiv \delta p/\lambda$, and 
\begin{align}
\nonumber
N=&\frac{1+\cos{(2\lambda\bar{q})}e^{-1/2\delta^2}}{2}\alpha_0^w
+\sin{(2\lambda\bar{q})}e^{-1/2\delta^2}\mathrm{Im}(\alpha_1^w)
\\
&+\frac{1-\cos{(2\lambda\bar{q})}e^{-1/2\delta^2}}{2}\alpha_{1,1}^w .
\end{align}
The exact moments are, for even $j$, 
\begin{align}
\langle p^j\rangle_f =\overline{p^j}+\frac{ \alpha^w_0+\alpha^w_{1,1}}{2N}{\sum_{k=1}^{j/2}}\binom{j}{2k} \overline{p^{j-2k}}
 \lambda^{2k} , 
\label{eq:exmomseven}
\end{align}
and, for odd $j$, 
\begin{align}
\langle p^j\rangle_f =\frac{1}{N}& 
\sum_{k=0}^{(j-1)/2} \binom{j}{2k} \overline{p^{2k}}
  \lambda^{j-2k} \mathrm{Re}(\alpha^w_1) .
\label{eq:exmomsodd}
\end{align}

For even $j$, the interpolation formulas are given by Eqs.~\eqref{eq:gaussmomseven} and 
\eqref{eq:gaussmomsodd}, with the $\alpha^w$ given above, or, equivalently, by the first non vanishing order expansion of Eqs.~\eqref{eq:exmomseven} 
and \eqref{eq:exmomsodd}
\begin{align}
\langle p^j\rangle_f \simeq \overline{p^j}+ \lambda^{2}\binom{j}{2} \overline{p^{j-2}} \frac{ \alpha^w_0+\alpha^w_{1,1}}{2N_2} 
 , 
\label{eq:intmomseven}
\end{align}
for even $j$, and, for odd $j$, 
\begin{align}
\langle p^j\rangle_f \simeq & 
\lambda j\overline{p^{j-1}} \frac{\mathrm{Re}(\alpha^w_1)}{N_2} .
\label{eq:intmomsodd}
\end{align}
As Figs. \ref{fig:comp} and \ref{fig:compmoms} show, 
the interpolation formulas work satisfactorily over the whole range of pre- 
and postselection. 
We notice that the validity of the interpolation is especially sensitive to the value of $\lambda \bar{q}$, 
since it contributes to first order. 
 
\begin{figure}[t!]
\includegraphics[width=2.4in]{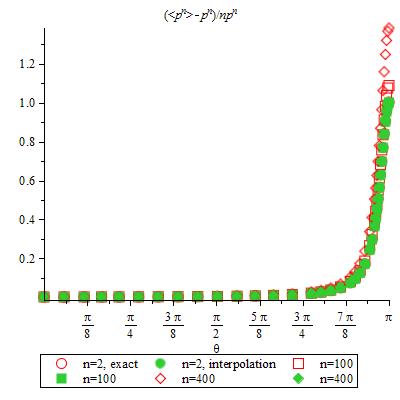}
\caption{
\label{fig:scaledmoms}
Universal scaling for the moments.}
\end{figure}
We notice that the even moments, to lowest order in $\lambda$, present a universal scaling
\begin{equation}
\frac{\langle p^j\rangle_f -\overline{p^j}}{j\overline{p^j}}\simeq  \frac{\alpha^w_0+\alpha^w_{1,1}}{4\Delta^2 N}. 
\end{equation}
Since for NOPPS $\alpha^w_{1,1}\gg \alpha^w_0$ and $N\simeq (1/2\delta)^2 \alpha^w_{1,1}$, all renormalized even moments take the value $\delta p^2/\Delta P^2$ 
for orthogonal pre- and post-selection, as shown in Fig. \ref{fig:scaledmoms}, where we compare $j=2,100,400$, showing that the scaling breaks down at 
$n^*\simeq (\Delta P/\lambda)^2$.  
Odd moments, instead, vanish for orthogonal pre- and post-selection, since the second order in $\lambda$ in the numerator is zero. In principle, one 
should retain the third, or higher order terms, but this would give a  value $o(\lambda/\Delta P)$  for $\langle p^j\rangle_f$, hence the interpolation is still satisfactory.

\section{Conclusions}
In conclusion, we have showed that the statistics of the weak measurement 
is interpolated satisfactorily by the rational function $f/g$, where both $f$ and $g$ are second 
order polynomials in the interaction strength $\lambda$. 
The coefficients 
are written in terms of averages 
 over the initial state of the probe, and of 
a complex number, the canonical weak value, 
$A^w=\Tr{\{\Hat{\rho}_f\Hat{A}\Hat{\rho}_i\}}/\Tr{\{\Hat{\rho}_f\Hat{\rho}_i\}}$, 
plus a real number $B^w=\Tr{\{\Hat{\rho}_f\Hat{A}\Hat{\rho}_i\Hat{A}\}}/\Tr{\{\Hat{\rho}_f\Hat{\rho}_i\}}$. 
In some cases, the coefficients of $\mathrm{Re}(A^w)$ and $\mathrm{Im}(A^w)$  
may vanish in the numerator $f$, so that it is necessary to introduce another complex number, 
$C^w=\Tr{\{\Hat{\rho}_f\Hat{A^2}\Hat{\rho}_i\}}/\Tr{\{\Hat{\rho}_f\Hat{\rho}_i\}}$,  
in order to obtain a satisfactory interpolation. 

The results presented here apply to an instantaneous interaction, so that the Hamiltonian evolution 
of the detector can be neglected. If this is not the case, the coefficients of the expansion have more involved 
expressions, and will be considered in a forthcoming work \cite{DiLorenzo2011i}. 
	
This work was supported by Funda\c{c}\~{a}o de Amparo \`{a} Pesquisa do 
Estado de Minas Gerais through Process No. APQ-02804-10.

\end{document}